\documentclass[12pt,thmsa]{article}
\usepackage{sw20aip}

\input{tcilatex}
\begin{document}

\author{Emilio Santos \and Departamento de F\'{i}sica. Universidad de Cantabria}
\title{Bell\'{}s theorem and the experiments: Increasing empirical support to local
realism?}
\date{October, 16, 2004 }
\maketitle

\begin{abstract}
It is argued that local realism is a fundamental principle, which might be
rejected only if experiments clearly show that it is untenable. A critical
review is presented of the derivations of Bell\'{}s inequalities and the
performed experiments, with the conclusion that no valid, loophole-free,
test exists of local realism vs. quantum mechanics. It is pointed out that,
without any essential modification, quantum mechanics might be compatible
with local realism. This suggests that the principle may be respected by
nature.
\end{abstract}

\section{Introduction}

Forty years have elapsed since John Bell\cite{Bell} , \cite{speakable}
discovered his celebrated inequalities. These inequalities, which involve
measurable quantities, provide necessary conditions for \textit{local realism%
}. Bell also proved that, in some experiments with ideal set-ups, the
predictions of quantum mechanics violate the inequalities. During these four
decades a lot of papers have been written pointing out quantum-theoretical
violations of the inequalities in very many different phenomena, but only a
few dozens empirical tests have been actually performed. The results of all
performed experiments are compatible with local realism and, with few
exceptions, agree with the quantum predictions (see Secs. 5, 6 and 8 below).
The logical interpretation of these facts, unbiased by theoretical
prejudices, should be that there is no empirical evidence against local
realism and that quantum mechanics has been confirmed, the few disagreements
with its predictions being of little significance. Nevertheless the standard
wisdom is that local realism has been refuted, which is concluded because
allegedly plausible extrapolations of the empirical results could violate a
Bell inequality.

In my view the current wisdom is misleading and harmful for the progress of
science. Misleading because it attempts answering a fundamental scientific
question by means of a subjective assessment of plausibility. Harmful
because it discourages people from making the necessary effort to perform a
real, loophole-free, test.

The long time elapsed without a true disproof of local realism may be
compared, for instance, with the discovery of parity non-conservation, which
required a few months to go from the theoretical paper by Lee-Yang, in 1957,
to the uncontroversial (loophole-free) experiment by Wu et al. I think that
the logical conclusion of the long standing unsuccessful effort to disprove
local realism is that it is preserved by nature.

In this paper I shall begin analyzing the concept of local realism and its
relevance in physics (Sec. 2). Then I shall sketch the derivation of
Bell\'{}s inequalities, distinguishing those which follow just from local
realism (Sec. 3) from those which require auxiliary assumptions (Sec. 4).
After that I shall review the experiments aimed at testing local realism vs.
quantum mechanics (Secs. 5 and 6). Then I shall rebute the standard wisdom
that quantum mechanics is incompatible with quantum mechanics (Sec.7).
Finally, after a digression on philosophy and sociology of science (Sec. 8)
I shall discuss the consequences to be drawn from the performed experiments
(Sec. 9).

\section{Local realism and its relevance in natural science}

It is not easy to define \textit{realism }with a few words\textit{,} as is
proved by the existence of whole books devoted to the subject. Here I shall
give a simple definition appropriate for physics. \textit{Realism is the
belief that material bodies have properties independent of any observation,
and that the results of any possible measurement depend on these properties.}
The said properties may be called ``elements of reality''\cite{EPR} and are
frequently identified with \textit{hidden variables}\cite{Bell2} , \cite
{Mermin} . However I think that the latter correspond rather to the
parameters used for the description of the said properties and should not be
confused with the former.

Realism alone, as defined above, does not contradict quantum mechanics. In
order to clarify the point I shall give an example. If I throw upwards a
coin, after a while the coin will collide with, say, a table and will soon
become at rest on it, with either the head or the tail upwards. The
described experiment consists, as is typical, of the \textit{preparation} of
the state of a system (the coin thrown upwards) followed by the \textit{%
evolution} of the system and finishing by the \textit{measurement} of a
quantity on it. Our intuition says that the result (head or tail) is
determined by the elements of reality of the coin during the fly. Or maybe,
taking into account the unavoidable existence of non-idealties (e.g.
friction with the air), the elements of reality just determine the
probability of the result. In any case we should carefully distinguish
between the \textit{observable} (head or tail) and the \textit{elements of
reality (}associated to motion of the coin\textit{)}. The relevant lesson of
our example is that the result of a measurment depends on both the measured
system (the coin) and the measuring apparatus (the table). Sometimes the 
\textit{observable }(head-tail in our example) is even devoid of sense
without the \textit{measuring apparatus} (the table). Therefore it is not so
strange that quantum mechanics forbids the ``simultaneous existence of
definite values for some observables'', namely those which cannot be
measured toghether. This is the essential content of the Kochen-Specker
theorem forbidding non-contextual hidden variables\cite{Mermin}. Our example
shows that the validity of the theorem does not preclude realism.

Some quantum physicists may consider that realism is just a philosophical
opinion which may or may not be true, but I disagree. In my view natural
science would be impossible without accepting realism as defined above.
Actually, even the most pragmatic quantum physicists would admit that states
of physical systems have some ``capabilities'' of influencing the results of
eventual future measurements on the system. It is a rather semantic question
whether we name these capabilities ``elements of reality''.

\textit{Locality is the belief that no\ influence may be transmitted with a
speed greater than that of light. }Thus we might identify \textit{locality}
with\textit{\ relativistic causality}. The concept of locality is subtle,
however. In fact, quantum mechanics is local in the sense that it forbids
the transmission of superluminal signals (say from a human being to another
one), but local realism as analyzed here is stronger than that. At a
difference with the idea of realism which I consider as an unvoidable
requirement for the existence of science, locality derives from our
experience at the macroscopic level and might be violated without
demolishing the whole building of physics. That is, we might assume that
some influences travel at a speed greater than that of light even if this
fact does not allow the transmission of superluminal signals. This seemed
the position of John Bell\cite{Bell4} .

In spite of this I think that locality is also important, that is local
realism is so fundamental a principle of physics that it should not be
rejected without \textit{extremely strong reasons}, an opinion which I
believe is quite close to what Einstein mantained until his death\cite
{Einstein} . On the other hand the question of \textit{local hidden variables%
} is less relevant than the question of local realism. It is true that if
local realism is untrue local hidden variables would be impossible, but if
local realism is true local hidden variables may still be useless in
practice, although possible in principle. Thus I shall refer to local
realism, rather than to local hidden variables, in the rest of this article.

\section{The Bell inequalities}

From what we have said it might appear that local realism is a purely
philosophical concept. But a \textit{physical} necessary condition for 
\textit{local realism} was introduced by John Bell\cite{Bell75} as follows: 
\textit{Any correlation between measurements performed at different places
should derive from events which happened in the intersection of the past
light cones of the measurements}. In order to give an empirical content to
the statement Bell considered a generic experiment consisting of the
preparation of a pair of particles (or, more generally, physical systems)
which are let to evolve in such a way that the two particles go to
macroscopically distant regions ( the argument that follows has been exposed
in more detail elsewhere\cite{FS}.) Thus Bell searched for the probability,
p(A,a;B,b), of getting the result \textit{a} in the measurement of an
observable A of the first particle and the result \textit{b} in the
measurement of the observable B of the second particle. He proposed that, if
local realism holds true, the probability could be written

\begin{equation}
p(A,a;B,b)=\int \rho (\lambda )P_{1}(\lambda ;A,a)P_{2}(\lambda
;B,b)d\lambda ,  \label{1}
\end{equation}
where $\lambda $ is one or several parameters which contain all relevant
information about the intersection of the past light cones of the two
measurements. An expression similar to $\left( \ref{1}\right) $ for the
total probability $p(A,a),$ of getting the result\textit{\ a} in the
measurement of the observable A on the first particle, follows at once from
the fact that it is unity the sum of probabilities associated to particle 2.
That is 
\begin{equation}
p(A,a)=\sum_{b}\int \rho (\lambda )P_{1}(\lambda ;A,a)P_{2}(\lambda
;B,b)d\lambda =\int \rho (\lambda )P_{1}(\lambda ;A,a)d\lambda .  \label{1a}
\end{equation}
From now on we shall consider only dichotomic observables, so that the
result of the measurement of the observable A may be only 1 (yes) or 0
(not). Thus we shall simplify the notation writing $P_{1}(\lambda ,A)$ (or $%
P_{2}(\lambda ,B)$ ) for $P_{1}(\lambda ;A,a)$ (or $P_{2}(\lambda ;B,b))$,
and p(A,B) ( or p(A)) for the left side of $\left( \ref{1}\right) $ ( $%
\left( \ref{1a}\right) )$ so that eqs.$\left( \ref{1}\right) $ and $\left( 
\ref{1a}\right) $ will be written 
\begin{equation}
p(A)=\int \rho (\lambda )P_{1}(\lambda ,A)d\lambda ,\;p(A,B)=\int \rho
(\lambda )P_{1}(\lambda ,A)P_{2}(\lambda ,B)d\lambda .  \label{1c}
\end{equation}

The functions $P$ and $\rho $ in the formula fulfil the conditions required
for probabilities and probability densities, respectively. That is 
\begin{equation}
\rho (\lambda )\geq 0,\int \rho (\lambda )d\lambda =1,  \label{2}
\end{equation}
\smallskip 
\begin{equation}
P_{1}(\lambda ,A),P_{2}(\lambda ,B)\geq 0,  \label{3}
\end{equation}
\smallskip 
\begin{equation}
P_{1}(\lambda ,A),P_{2}(\lambda ,B)\leq 1.  \label{4}
\end{equation}
It is important to stress that the value of $P_{1}(\lambda ,A)$ is assumed
to be independent of B, that is independent on what measurement is performed
on the second particle, which is Bell\'{}s condition of locality. This
independence has been called ``parameter independence'', which is compatible
with a possible ``outcome dependence'', that is the results of the
measurements of A and B may be correlated\cite{Jarret}. Hence, using the
notation A\'{}, B \'{} for the result 0 in the measurement of A and B
respectively, we obtain a similar independence for the measurable
probabilities 
\[
p(A)=p(A,B)+p(A,B^{\prime })=p(A,D)+p(A,D^{\prime })=... 
\]
Parameter independence holds true also in quantum mechanics and it
guarantees that superluminal communication is not possible.

\smallskip From the conditions $\left( \ref{1c}\right) $ to $\left( \ref{4}%
\right) $ it is possible to derive inequalities involving only measurable
probabilities. We consider an experiment in which we prepare once and again,
say 4N times (N $\gg $ 1), a pair of particles in a given state, the same
for all preparations. Here, \textit{the same} means that the parameters
which may be controlled in the preparation have the same values. After N
preparations, chosen at random amongst the 4N made, we measure the
dichotomic observables A and B of the two particles. After another N
preparations, also chosen at random, we measure the dichotomic observables C
and D. Similarly C with B are measured N times, and A with D also N times.
We assume that the result of the measurement of any of the observables may
be either 0 or 1, and call p(A, B) the probability of getting the result 1
for both observables, A and B (the frequencies measured in the experiment
should approach the probabilities if N is large enough). Similarly we may
define the probabilities p(A, D), p(C, B) and p(C, D), and also the
probability p(A) corresponds to getting the value 1 in the measurement of A
and any value (1 or 0) in the measurement of B, or D, performed on the
partner particle, and similar for p(B). It is an easy task to derive, from$%
\left( \ref{1}\right) $ to $\left( \ref{4}\right) ,$ inequalities involving
measurable probabilities. For instance\cite{CH} 
\begin{equation}
p(A,B)+p(A,D)+p(C,B)-p(C,D)\leq p(A)+p(B).  \label{CH}
\end{equation}

This inequality may be related to the existence of a ``metric'' in the set
of propositions associated to the results ``yes'', ``no'' in the four
measurements. In fact we may define a formal (not measurable) joint
probability distribution on the observables \{A, B, C, D\} by means of
expressions similar to $\left( \ref{1c}\right) $ applied to the four
observables, the six pairs $\left\{ AB,AC,AD,BC,BD,CD\right\} $ and the four
triples $\left\{ ABC,ABD,ACD,BCD\right\} ,$ in spite of some of them not
being actually measurable (e. g. p(A,C) cannot be got empirically because A
and C correspond to alternative, incompatible, measurements on the same
particle). Now the mere possibility of defining a formal joint probability
implies the existence of a metric in the set of propositions (yes-no
experiments) and the essential property of the metric is the fulfillement of
triangle inequalities, which are closely related to the inequality $\left( 
\ref{CH}\right) $. But I shall not pursue the subject here (details may be
seen elsewhere\cite{Santos3}.)

\section{Bell\'{}s vs. tested inequalities. The CHSH case.}

Soon after Bell\'{}s discovery\cite{Bell} in 1964, it was realized that no
performed experiment had shown a violation of local realism. Furthermore, no
simple experiment could do the job. In my view, the difficulty is a proof
that it is wrong the wisdom according to which quantum mechanics predicts
``highly non-local effects''. The truth is that non-local effects, if any,
are extremely weak and difficult to observe.

In 1969 Clauser, Horne, Shimony and Holt (CHSH)\cite{CHSH} made the first
serious proposal for an empirical test of Bell\'{}s inequality . They
suggested the measurement of the polarization correlation of optical photon
pairs. By optical we mean that the corresponding frequencies are in the
visible, the near ultraviolet or the near infrared parts of the spectrum.
The mentioned authors derived the Bell inequality 
\begin{equation}
S\equiv E(A,B)+E(A,D)+E(C,B)-E(C,D)\leq 2,  \label{CHSH}
\end{equation}
where \{A, C\} correspond to two possible positions of a polarization
analyzer for the first photon and \{ B, D\} for the second. The correlations
are defined by 
\begin{equation}
E(X,Y)=p_{++}(X,Y)+p_{--}(X,Y)-p_{+-}(X,Y)-p_{-+}(X,Y),  \label{corr}
\end{equation}
with X = A or C, Y\ = B or D, p$_{++}(X,Y)$ being the probability that the
polarization of the first photon is found in the plane X, and that of the
second in the plane Y, p$_{+-}(X,Y)$ the probability that the polarization
of the first photon is found in the plane X and that of the second is in the
plane perpendicular to B, etc.

It is not difficult to see that the $\left( \ref{CHSH}\right) $ inequality
is equivalent to $\left( \ref{CH}\right) $ provided that the sum of the four
probabilities involved is unity, that is 
\begin{equation}
p_{++}(X,Y)+p_{--}(X,Y)+p_{+-}(X,Y)+p_{-+}(X,Y)=1.  \label{norm}
\end{equation}
In fact in this case it is easy to go from $\left( \ref{CHSH}\right) $ to $%
\left( \ref{CH}\right) ,$ or viceversa, by repeated use of relations like 
\begin{eqnarray}
p_{++}(X,Y) &=&p(X,Y),\;p_{+-}(X,Y)=p(X)-p(X,Y),  \nonumber \\
\;p_{--}(X,Y) &=&1-p(Y)-p_{+-}(X,Y).  \label{rel}
\end{eqnarray}
With respect to the empirical tests, however, the two inequalities look
rather different, and only the inequality $\left( \ref{CH}\right) $ may be
easily adjusted to actual experiments. In fact, in the experiments either eq.%
$\left( \ref{norm}\right) $ is not true, thus $\left( \ref{CHSH}\right) $
not being a true Bell inequality (it cannot be derived from local realism
alone) or the quantities E(X,Y) are no longer correlations, as we explain in
the following.

In typical experiments there are two arms in the apparatus, each one
consisting of a lens system followed by a polarization analyzer (polarizer,
for short) and a detector (for the moment we do not consider the case of
two-channel analyzers, but see below). Thus we may interpret p(X,Y) as the
probability that both photons are detected, after crossing the appropriate
polarizers, and p(X) the probability that the ``red'' photon of the pair is
detected, with independence of what happens to the ``green'' photon (for
clarity of exposition we attach fictitious colours, red and green, to the
photons of a pair). However, if this interpretation is carried upon the
quantities E(X,Y), via the relations $\left( \ref{rel}\right) ,$ such
quantities would be correlations only in the case that both photons of every
pair arrive at the polarizers and every photon is detected (with 100\%
efficiency) whenever it has crossed the corresponding polarizer. But this
idealized situation never happens.

The current practice in recent experiments is to use two-channel polarizers,
with a detector after each outgoing channel. Attaching the labels + or - to
the detectors after the first or second outgoing channel of a polarizer,
respectively, it is possible to define p$_{++}$ as the probability that both
photons are detected in detectors with label +, p$_{+-}$ the probability
that the red photon is detected in a detector with label + and the green
photon in a detector with label -, etc. With this interpretation the
quantities E(X,Y) of $\left( \ref{corr}\right) $ are indeed true
correlations and the inequality $\left( \ref{CHSH}\right) $ is never
violated in actual experiments, because all probabilities p$_{++}$, p$_{+-}$%
, etc. are much smaller than unity due to the low collection-detection
efficiency (i. e. for most photon pairs only one photon, or none, is
detected). The ``solution'' proposed for this problem has been to
renormalize the probabilities defining the correlations by 
\begin{equation}
E^{*}(X,Y)=\frac{p_{++}(X,Y)+p_{--}(X,Y)-p_{+-}(X,Y)-p_{-+}(X,Y)}{%
p_{++}(X,Y)+p_{--}(X,Y)+p_{+-}(X,Y)+p_{-+}(X,Y)}.  \label{corr1}
\end{equation}
Thus people use the inequality (compare with $\left( \ref{CHSH}\right) $) 
\begin{equation}
S^{*}\equiv E^{*}(A,B)+E^{*}(A,D)+E^{*}(C,B)-E^{*}(C,D)\leq 2,  \label{CHSH1}
\end{equation}
in the empirical tests. Indeed, this is the inequality violated in most of
the recent experiments. The inequality, however, cannot be derived from eqs.$%
\left( \ref{1c}\right) $ to $\left( \ref{4}\right) $ alone (without
additional assumptions) and therefore it is not a genuine Bell inequality.

\section{Experiments using optical photons}

The first experimental test using optical photons was made by Freedman and
Clauser\cite{Freedman}. They used photon pairs produced in the decay of
excited calcium atoms via a 0-1-0 cascade. That is, the initial and final
atomic states had 0 total angular momentum, so that the two emitted photons
were entangled in polarization. The dichotomic observables measured were
detection or non-detection of a photon, after it passed through a polarizer
. The labels A and C are associated to two different positions of the
polarizer for the ``red'' photon and similarly B and D for the ``green''
one. The authors were aware that the inequality $\left( \ref{CH}\right) $
could not be violated with the technology of the moment because the
detection efficiencies of the available detectors were too small (less than
10\%.) As the left hand side of the inequality $\left( \ref{CH}\right) $ is
proportional to the efficiency squared, whilst the right side is
proportional to the efficiency, the latter is more than ten times the
former, so that the inequality is very well fulfilled.

More specifically, the predition of quantum mechanics for the experiment may
be summarized as follows, with some simplifications for the sake of clarity.
The measurable quantities in the experiment are the single rates, R$_{1}$
and R$_{2}$, and coincidence rate, R$_{12}$($\phi ),$ the latter being a
function of the angle, $\phi ,$ between the polarizer\'{}s planes X and Y.
In terms of the production rate, R$_{0}$, in the source they are given by 
\begin{equation}
R_{1}(A)=R_{2}(B)=\frac{1}{2}R_{0}\eta ,\;R_{12}(X,Y)=\frac{1}{4}R_{0}\eta
^{2}\alpha (1+V\cos \left( 2\phi \right) ).  \label{pred}
\end{equation}
Here $\alpha $ is an angular correlation parameter and $\eta $ is the
overall detection efficiency of a photon, which includes collection
efficiency and quantum efficiency of the detectors (for simplicity we put
the same efficiency $\eta $ for the red and the green photons, which is
approximately true in practice, but the generalization would be rather
trivial). In actual experiments the quantum prediction $\left( \ref{pred}%
\right) $ is confirmed, except for small deviations which are not considered
significant.

The probabilities needed to test the inequality $\left( \ref{CH}\right) $
are just the ratios 
\[
p(A)=\frac{R_{1}}{R_{0}},\;p(B)=\frac{R_{2}}{R_{0}},\;p(X,Y)=\frac{%
R_{12}(\phi )}{R_{0}}. 
\]
The production rate, R$_{0}$, is not measured but it is not difficult to
show that, if we insert $\left( \ref{pred}\right) $ into $\left( \ref{CH}%
\right) ,$ R$_{0\text{ }}$ cancels out and the inequality becomes 
\[
\alpha \eta \left[ 1+\frac{1}{2}V\left( \sum_{1}^{3}\cos (2\phi _{j})-\cos
\left( 2\phi _{4}\right) \right) \right] \leq 2, 
\]
where $\{\phi _{j}\}$ are the angles between the polarization planes of the
analyzers, that is between A and B, A and D, C and B, C and D, respectively.
These angles fulfil $\phi _{1}+\phi _{4}=\phi _{2}+\phi _{3}$ and the
maximum of $\sum_{1}^{3}\cos (2\phi _{j})-\cos \left( 2\phi _{4}\right) $
with that constraint is 2$\sqrt{2}.$ Thus the Bell inequality $\left( \ref
{BI}\right) $ holds true, for any choice of polarizers positions, whenever 
\begin{equation}
\alpha \eta \left( 1+\sqrt{2}V\right) \leq 2.  \label{BI}
\end{equation}
In the actual experiment\cite{Freedman} V $\simeq $ 0.85, but $\eta $ $%
\simeq $0.0001, and $\alpha \simeq 1,$ so that the inequality was safely
fulfilled ($\eta $ is the product of the quantum efficiency, $\zeta ,$ of a
detector times the collection efficiency of the apertures, see below eq.$%
\left( \ref{coll}\right) ).$

Freedman and Clauser\cite{Freedman} found a ``solution'', to circumvent the
problem of the low detection efficiency, consisting of the replacement of
condition $\left( \ref{4}\right) $ by another one, called
``no-enhancement'', which they claimed \textit{plausible}. This assumption
states that, for any value of the parameter $\lambda ,$ the following
inequality holds true:

\begin{equation}
P_{1}(\lambda ,A)\leq P_{1}(\lambda ,\infty ),\;P_{2}(\lambda ,B)\leq
P_{2}(\lambda ,\infty )  \label{4a}
\end{equation}
where $P_{j}(\lambda ,\infty )$ are the probabilities of detection of the
photon with the corresponding polarizer removed. From inequalities $\left( 
\ref{1}\right) $ to $\left( \ref{3}\right) $ plus $\left( \ref{4a}\right) $,
the authors\cite{Freedman} derived the inequality 
\begin{equation}
p(A,B)+p(A,D)+p(C,B)-p(C,D)\leq p(A,\infty )+p(\infty ,B),  \label{FC1}
\end{equation}
where $p(A,\infty )$ ( $p(\infty ,B)$ ) is the probability of coincidence
detection with the polarizer corresponding to the red (green) photon
removed. The results of the measurement, and the quantum predictions, for
these probabilities are 
\[
p(A,\infty )=p(\infty ,B)=\frac{1}{2}\alpha \eta ^{2}, 
\]
and the inequality $\left( \ref{FC1}\right) $ implies 
\begin{equation}
\left( 1+\sqrt{2}V\right) \leq 2\Leftrightarrow V\leq \sqrt{2}/2,  \label{FC}
\end{equation}
to be compared with $\left( \ref{BI}\right) .$ This was the inequality
tested, and violated, in the commented experiment.

Note that, in sharp contrast with the obvious inequality $\left( \ref{4}%
\right) ,$ the inequality $\left( \ref{4a}\right) $ is not only empirically
untestable, it is \textit{counterfactual. }In fact, as said above, $\lambda $
is a set of parameters which contains all relevant information about the
intersection of the past light cones of the measurements. But the past light
cone of one measurement (with a polarizer in place) is necessarily different
from the past light cone of a different measurement (with the polarizer
removed). In order to give a meaning to the inequality $\left( \ref{4a}%
\right) $ it is necessary to compare a fact (one of the measurements) with a
belief (about what would have happened in a different experiment having the
same past light cone). For this reason I say that the inequality is
counterfactual. Of course, it may be checked empirically that \textit{the
average} over $\lambda $ of the left hand side is not greater than the
average of the right hand side, that is for any light beam the detection
rate does not increase when we insert a polarizer. (However, it might
increase if we insert a polarization rotator plus a polarization analyzer
when the incoming light is linearly polarized ). In summary, the first
alleged empirical disproof of local realism rests upon a counterfactual
belief qualified as plausible. Therefore, strictly speaking, it did not test
local realism. However I do not mean that the experiment was useless because
it opened an important new line of experimental research.

In the decade that followed the commented experiment, several similar
atomic-cascade experiments were performed\cite{CS} , \cite{Selleri6} . In
addition to the requirement of introducing untestable auxiliary assumptions
(like $\left( \ref{4a}\right) ),$ all of them had the problem of being
static. That is, the positions of the polarizers were fixed well before the
detection events took place. Therefore the experiments could not test
locality, in the sense of relativistic causality. In order to solve the
problem, Alain Aspect and coworkers\cite{Aspect} performed in 1982 a new
atomic-cascade experiment where (in some sense) the polarizers positions
were chosen when the photons were already in fly. However the inequality
tested was of the type $\left( \ref{FC1}\right) $ rather than a genuine Bell
inequality like $\left( \ref{CH}\right) .$

The experiment of Aspect is usully presented as the definite refutation of
local realism. One of the reasons is that, during the preparation of the
experiment, Aspect was in close contact with Bell, who approved it. Although
Bell was aware that there existed a loophole due to the low efficiency of
the available photon detectors, he considered acceptable to make a \textit{%
fair sampling assumption}. That is, to extrapolate the results actually got
in the experiment, with low efficiency detectors, to detectors 100\%
efficient. This amounts to testing an inequality obtained from $\left( \ref
{CH}\right) $ by dividing the right hand side by the efficiency, $\eta ,$
and the left side by $\eta ^{2}.$ The inequality so obtained is practically
the same as $\left( \ref{FC1}\right) .$ The fair sampling assumption was
justified by Bell\cite{Bell3} with the frequently quoted sentence: ``It is
hard for me to believe that quantum mechanics works so nicely for
inefficient practical set-ups and is yet going to fail badly when sufficient
refinements are made.'' But this sentence cannot be applied to the commented
experiments because \textit{the predictions of quantum mechanics for any
atomic-cascade experiment are compatible with local realism even if the
experiment is made with ideal set-up}, in particular 100\% efficiency
detectors, as is shown in the following\cite{Santos91}. Apparently Bell was
not aware of this fact before he untimely died in october 1990.

The atomic cascade decay, giving rise to a photon pair, is a three-body
problem with the consequence that the angle, $\chi ,$ between the directions
of emission of the two photons is almost uniformily distributed over the
sphere. This implies that the angular correlation parameter $\alpha $ (see $%
\left( \ref{pred}\right) )$ is almost independent of the angle $\chi $, that
is $\alpha \left( \chi \right) \simeq 1$. On the other hand both the overall
detection efficiency, $\eta ,$ and the ``visibility'', V, of the coincidence
curve are functions of the angle, $\theta ,$ determined by the apertures of
the lens system (as seen from the source). The dependence V($\theta )$ is a
loss of polarization correlation when the ``red'' and ``green'' photons do
not have opposite wavevectors. In the Aspect experiment, as in other
atomic-cascade experiments, the predicted functions are\cite{CS}

\begin{equation}
\eta =\frac{1}{2}(1-\cos \theta )\zeta ,\;V=1-\frac{2}{3}(1-\cos \theta
)^{2},\;\alpha \simeq 1,  \label{coll}
\end{equation}
where $\zeta $ is the quantum efficiency of the detectors. Using these
expressions it is easy to see that the maximum value of the left hand side
of $\left( \ref{BI}\right) $ is about 0.74 $\zeta ,$ and the inequality is
safelly fulfilled even for ideal detectors (i. e. $\zeta =1).$ The figure
should be multiplied times 2, giving 1.48 $\zeta ,$ if we assume that both
photons , red and green, may be detected in either detector. But still the
inequality $\left( \ref{BI}\right) $ holds true for any $\zeta $ ($\leq 1).$
In summary taking into account the low angular correlation of the photon
pairs produced in atomic cascades,\textit{\ these experiments cannot
discriminate between local realism and quantum mechanics. }In spite of this
fact, the Aspect experiment is quoted everywhere as the definite refutation
of local realism.

The problem of the lack of angular correlation might be solved if the recoil
atom were detected\cite{Susana} but that experiment would be extremely
difficult. A more simple solution is to use optical photon pairs produced in
the process of parametric down conversion, and this has been the source
common in all experiments since about 1984. See, for instance, the paper by
Kurtsiefer et al.\cite{Kurt} and references therein. At a difference with
atomic-cascade experiments, here the photons have a good angular
correlation. In fact the parameter $\alpha $ of $\left( \ref{pred}\right) $
as a function of the angle, $\chi ,$ between the wavevectors of the two
photons is such that the probability of detection of the green photon
conditional to the detection of the red one is just the quantum efficiency $%
\zeta $ (or close to it.) Thus putting the detectors in appropriate places
we may rewrite $\left( \ref{BI}\right) $ with $\zeta $ substituted for $%
\alpha \eta $, that is 
\begin{equation}
\zeta \left( 1+\sqrt{2}V\right) \leq 2.  \label{BI1}
\end{equation}
This inequality might be violated if V is close to 1 (which is achievable in
actual experiments) and $\zeta >2\left( \sqrt{2}-1\right) \simeq 0.82$. But
such a high value of the detection efficiency has not yet been achieved and
the low efficiency of detectors remains as a persistent loophole for the
disproof of local realism.

This difficulty has led to the use of the modified CHSH inequality $\left( 
\ref{CHSH1}\right) $ as the standard inequality tested in practically all
recent experiments with optical photons. These experiments use two-channel
polarizers, and the prediction of quantum mechanics for them may be
summarized in terms of four coincidence detection rates as follows 
\begin{eqnarray}
R_{++}\left( \phi \right) &=&R_{--}\left( \phi \right) =\frac{1}{2}\eta
R_{0}\left[ 1+V\cos \left( 2\phi \right) \right] ,\;  \nonumber \\
R_{+-}\left( \phi \right) &=&R_{-+}\left( \phi \right) =R_{++}\left( \phi +%
\frac{\pi }{2}\right) ,  \label{pred1}
\end{eqnarray}
whilst the single rates are usually not measured (or, at least, not reported
as relevant). In the actual experiments there are small departures from $%
\left( \ref{pred1}\right) $ which are not considered significant, but may be
relevant for the reasons (see end of this section.) As said above the
inequality tested is $\left( \ref{CHSH1}\right) ,$ and the probabilities
involved may be obtained from $\left( \ref{pred1}\right) $ as ratios between
the measured coincidence rates and the production rate. That is, putting $%
\left( \ref{pred1}\right) $ into $\left( \ref{corr1}\right) $ we get 
\begin{equation}
E^{*}=V\cos \left( 2\phi \right) .  \label{corr2}
\end{equation}
If this is used in $\left( \ref{CHSH1}\right) ,$ steps similar to those
leading to $\left( \ref{FC}\right) $ give 
\begin{equation}
S^{*}=2\sqrt{2}V\leq 2\Leftrightarrow V\leq \sqrt{2}/2.  \label{CS}
\end{equation}
This inequality looks the same as $\left( \ref{FC}\right) $, but here V is
obtained from measurements using two-channel polarizers. In practice V may
be got by at least three different procedures:

1) From the best fit of the measured correlation, $E(\phi ),$ to the
theoretical curve $\left( \ref{corr}\right) $, where the probabilities, $%
p_{++}(\phi )$, etc., are the ratios of measured rates, $R_{++}(\phi ),$
etc., to the production rate, R$_{0}$. It is easy to see that the fitting
does not require the measurement of R$_{0}$. We shall label just V the
quantity so obtained.

2) As half the ``visibility'' of the empirical curve $E\left( \phi \right) ,$
that is the difference between the maximum and the minimum values divided by
the sum. Again the value of R$_{0}$ is not required. I shall label V$_{A}$
this quantity.

3) From the value of $S^{*}$ measured for the angles $\phi _{1}=\phi
_{2}=\phi _{3}=\pi /8,\phi _{4}=3\pi /8,$ using the first equality $\left( 
\ref{CS}\right) .$ These angles provide the maximum value of $S^{*}$ if the
empirical data agree with $\left( \ref{pred1}\right) .$ This value will be
labelled V$_{B}$ and it is the quantity commonly used in the test of the
inequality $\left( \ref{CS}\right) $. Indeed in recent times it has become
standard practice to claim that local realism is refuted whenever V$_{B}$ 
\TEXTsymbol{>} $\sqrt{2}$/2.

According to quantum predictions the equality V = V$_{A}$ = V$_{B}$ should
hold true, but in actual experiments there are small differences between
them. On the other hand some natural families of local realistic models
predict inequalities involving the quantities V$_{A}$ and V$_{B}$ \cite
{Santos4}. One of these inequalities has already been tested with the result
that it is fulfilled, whilst the equality predicted by quantum mechanics
seems to be violated \cite{Genovese}.

A procedure to circumvent the low efficiency loophole in experiments with
optical photons has been proposed recently using homodyne detection instead
of photon counting\cite{Garcia}, \cite{Nha} . 

\section{Other experiments aimed at testing local realism}

In his pioneer work\cite{Bell} Bell used the example of an
Einstein-Podolsky-Rosen-Bohm\cite{Bohm1} system, that is a pair of spin-1/2
particles with zero total spin. Thus it is not strange that some experiments
have been proposed consisting of the measurement of the spin correlation of
two spin-1/2 particles. The use of massive particles has the advantage that
they may be quite reliably detected, so that such experiments do not suffer
from the detection loophole. The proposed experiments use non-relativistic
particles. As far as I know, no experiment of this kind has been proposed
using relativistic particles. The reason is probably the difficulty for
producing a pair with zero total spin if we take into account that spin of
relativistic particles is not stricly conserved (only the total angular
momentum of a free particle is strictly conserved in Dirac\'{}s theory).

The non-relativistic particles present the problem that it is difficult to
guarantee the space-like separation of the measurements. As an example, we
may consider the experiment proposed by Lo and Shimony\cite{Lo}. It consists
of the dissociation of molecules with two sodium atoms followed by the
measurement of their spins by means of a Stern-Gerlach apparatus. The
typical velocity of the sodium atoms, after dissociation, is about 3$000$
m/s and the length of the measuring magnets 0.25 m. this giving a
measurement time about 10$^{-4}s.$ Thus, in order that the measurements were
space-like separated, the Stern-Gerlach apparatused should be distant by
more than 30 km. It is rather obvious that such experiment could not be a
practical test of local realism as defined above. Similar problems appear in
the proposed experiment by Adelberger and Jones\cite{Jones} using neutron
pairs. The neutrons should collide at low energy in order to insure a pure
S-wave scattering so that, by Pauli\'{}s principle, the total spin should be
zero. Again the distance between the spin measurements (by scattering with
magnetized material) should be extremely large in order to be possible the
violation of locality (relativistic causality).

In addition there are fundamental constraints, derived from Heisenberg
uncertainty principle, on experiments using non-relativistic particles\cite
{Santos5}. For instance, let us assume that the particle detectors are
static and placed on opposite sides and at a distance L from the source
each, so that the distance between detectors is 2L. If the particles have
mass m (the same for both, for simplicity) and travel at a velocity v, then
the initial position and velocity are uncertain by, at least, $\Delta
x\Delta v=\hbar /2m.$ Thus the arrival time at the detectors will be
uncertain by, at least, $\sqrt{2\hbar L/(mv^{3})}.$ We may be sure that the
measurements are space-like separated only if this quantity is smaller than
2L/c, which leads to the constraint L$\geq 2\hbar c^{2}/(mv^{3}),$ a
macroscopic quantity (for instance, in the experiment proposed by Lo and
Shimony\cite{Lo} this gives about 1 m.) The quantity is not so big as to put
unsurmountable practical difficulties, but it shows that a Bell test using
non-relativistic particles requires measurements made at quite macroscopic
distances.

An experiment using the scattering of non-relativistic protons was performed
by Lamehi-Rachti and Mittig in 1976\cite{Lamehi}. The spin components of the
protons were measured by scattering on carbon foils. The experimental
results agreed with quantum predictions, but the auxiliary assumptions
needed for the experiment to be a test of a Bell inequality were stronger
than in experiments with optical photons. An experiment has been recently
performed\cite{Rowe} using two $^{9}$Be$^{+}$ ions in a trap, each of which
behaves as a two-state systems. It has been claimed, and widely commented,
that the experiment ``has closed the detection loophole'' because the atoms
may be detected with 100\% efficiency. However the distance between ions in
the trap, 3$\mu m,$ was very small. Although this distance is about 100
times the size of an ion wavepacket, it is 10$^{6}$ times smaller than the
wavelength of the photons involved in the atomic transitions between the two
levels (compare with the fundamental constraints commented in the previous
paragraph). In these conditions the experiment cannot test locality in the
sense of relativistic causality.

A loophole-free experiment involving spin measurements of atoms has also
been proposed. It consists of the disociation of mercury molecules followed
by the measurement of nuclear spin correlation of the atoms\cite{Fry}. In
order to make the measurment time very short, the idea was to use a
polarized pulse of laser light, which would induce selectively the
ionization of the atom when it is in one specific spin state (say up) but
not in the other possible state (down). After several years of preparation,
the detailed proposal of the experiment was published in 1995, but nine
years later no results have been reported. (In the Oviedo Conference, held
in July 2002, Fry reported that important difficulties had been found.
Fry\'{}s talk was not published in the proceedings\cite{Ferrero}).

Many other experimental tests of a Bell inequality have been performed or
proposed, each one suffering from loopholes. For instance, several
experiments have been performed measuring the polarization correlation of
gamma rays produced in the decay of positronium, one of the experiments
violating the quantum prediction\cite{Selleri6}. These experiments have the
difficulty that the polarization cannot be measured with high enough
precision.

There have been also proposals using high energy particles. For instance,
the strangeness oscillations of pairs $K^{0}-\overline{K}^{0}$ have been the
subject of many papers\cite{Selleri5} ,\cite{Bramon} , but no loophole-free
violation of a Bell inequality seems possible in this case due to the small
decay time of the short K$^{0}$ in comparison with the oscillation period.
Also an experiment has been recently performed using B$^{0}$ mesons\cite{Go}%
, but here also the damping made impossible the violation of a Bell
inequality, and only a normalization of the correlation function to the
undecayed pair leads to the violation of $\left( \ref{CHSH1}\right) $, not a
genuine Bell inequality.

In recent years a lot of effort has been devoted to the so-called ``tests
without inequalities''\cite{GHZ}. The idea is to prepare a system in some
state and perform a measurement such that the quantum prediction is definite
(say ``yes'') but the prediction of any local realistic model is the
opposite (``no''). For a proof of the incompatibility between local realism
and quantum mechanics, in ideal experiments, the proposal is very appealing
but from a practical point of view the possible experiments are less
reliable than those resting upon Bell\'{}s inequalities. In particular they
require an extreme control of the purity of the prepared state, which is not
the case in the Bell tests (see section 4). An experimental test of local
realism resting upon the idea has been performed\cite{Mandel}, but the
experiment is not conclusive, as is shown be the existence of a local model
reproducing the results\cite{Adan}.

In summary, no performed experiment has been able to test a genuine Bell
inequality with the condition that the measurements are performed at
space-like separation. And, as far as I know, only a detailed proposal for a
loophole-free experiment with available technology exists\cite{Fry}, but
this experiment seems to present unsurmountable difficulties. In consequence
local realism has not been refuted. Furthermore it is the case that,
strictly speaking, \textit{local realism has not yet been tested against
quantum mechanics. }That is no experiment has been performed able to
discriminate between local realism and quantum mechanics.

\section{Is quantum mechanics truly incompatible with local realism?}

The standard wisdom of the community of quantum physicists is that local
realism does not hold true in nature. Certainly this opinion does not follow
from just the results of the empirical tests of Bell\'{}s inequalities
because, as commented in the two previous sections, there are loopholes in
all performed experiments. Actually the current wisdom derives from the
theoretical argument that\textit{\ the validity of local realism would imply
that quantum mechanics is false (}Bell\'{}s theorem). And, for good reasons,
nobody is willing to accept that quantum mechanics is wrong. Thus the
scientific community dismisses the mentioned loopholes as irrelevant (see,
e. g., the relatively recent article by Lalo\"{e}\cite{Laloe}, excelent in
most other respects.) However \textit{a violation of local realism is no
more acceptable than a violation of quantum mechanics}, for the reasons
explained in section 2. Consequently there exists a real problem whose only
solution seems to me the compatibility of local realism with quantum
mechanics, or some ``small'' modification of this theory. But, is it
possible to modify quantum mechanics without destroying its formal beauty
and its impressive agreement with experiments?. In the following I argue for
this possibility.

According to the traditional formulation, quantum mechanics consists of two
quite different ingredients: the formalism (including the equations) and the
theory of measurement, both of which are postulated independently. (Actually
the two ingredients are to some extent contradictory, because the quantum
evolution is continuous and deterministic except during the measurement,
where the ``collapse of the wavefuction'' is discontinuous and stochastic.
Thus the modern approach tends to remove any \textit{postulated }theory of
measurement, see below). We must assume that the quantum equations are
correct, because the extremely accurate agreement between the predicted and
the measured, for instance in quantum electrodynamics, cannot be explained
otherwise. In contrast, only a small part of the quantum theory of
measurement is really used in most experiments, which suggests that it might
be substantially weakened. For instance, the postulate about position
measurements, i. e. Born\'{}s rule, is enough for the interpretation of all
scattering experiments.

The point is that standard proofs of ``Bell\'{}s theorem'' rest upon the
theory of measurement (and preparation of states). In fact, in a typical
proof it it assumed that: 1) A pure spin zero state of a system of two
spin-1/2 particles may be manufactured in the laboratory with the two
particles able to fly, mantaining the same joint spin state, up to
macroscopic distances, b) The spin projection of each particle, along any
freely chosen direction, may be measured with arbitrary small error. Both
these assumptions might be false without any danger for the formalism and
the basic equations (Dirac\'{}s, Maxwell\'{}s, etc.) of the theory.
Consequently I guess that a weakening of the standard measurement theory,
without touching the formalism, might make quantum mechanics compatible with
local realism. For instance, the weakening of the preparation and
measurement assumptions might be as follows. It is frequently assumed that
there is an one-to-one correspondence between the possible states of a given
physical system and the vectors in the Hilbert space, except for
superselection rules. That assumption is called \textit{superposition
principle}. But the unrestricted superposition principle also implies an
one-to-one correspondence between self-adjoint operators and observables.
This is because any self-adjoint operator may be written as a linear
combination of projectors onto its eigenstates (i. e. subspaces in the
Hilbert space) and each projector should be an observable if all states are
physically realizable and distinguishable. However the one-to-one
correspondence between self-adjoint operators and observables is far too
strong from a physical point of view because, how could we measure an
observable like $x^{m}p^{n}+p^{n}x^{m}$ with very large integers m and n?.
Thus it seems more appropriate to assume that \textit{only some}
self-adjoint operators represent observables and \textit{only some} vectors
represent states. We might go a step further and assume that \textit{only} 
\textit{some density matrices }represent physical realizable states. (For
instance we might restrict the states to density matrices fulfilling an
inequality like Tr$\left( \rho ^{2}\right) \leq k<1$ which, for some k,
might be sufficient to prevent the violation of Bell\'{}s inequalities\cite
{Santos7}. But I mention this possibility just as an illustrative example.)

If any proof of Bell\'{}s theorem resting upon the quantum theory of
measurement is invalid, a correct proof should involve a detailed study of
how to prepare the state and how to measure the observables able to violate
a Bell inequality. But the definite proof that some state and measurements
may be really made in the laboratory is to perform the actual experiment.
Thus I conclude that Bell\'{}s ``theorem'' cannot be proved by theoretical
arguments, i. e. it is not a theorem. It is just an argument \textit{%
suggesting} that some experiments might exist able to discriminate between
quantum and local realistic predictions. This conclusion does not mean a low
valuation of Bell\'{}s work, which I consider one of the most important
achievements in theoretical physics of the last 50 years. In any case Bell
never used the word ''theorem'' in this context, as far as I know.

In the modern approach, quantum measurement theory is not postulated but an
attempt is made at deriving it from the quantum formalism. I shall analyze
the results of this approach in the particular example of an optical photon
counter. I write ``counter'' in order to distinguish it from other types of
light detector. For instance, in astronomy a typical observational method is
to take a photographic plate of some region of the sky. In this case the
intensity of the light may be measured with a small error using a long time
of exposure, but no count of individual photons is made. I also include the
word ``optical'' because a single high energy photon (e. g. a gamma ray) has
a large enough energy to be detected with a probability close to 100\%. This
is not the case with optical photons. A counter of optical photons consists
of a macroscopic object (e. g. a piece of semiconductor) where there are
quantum systems (e. g. electrons) in metastable states. Typically the
metastable state appears because an external electric field is included
which, combined with the potential due to the ions, creates a potential well
where the electron is initially confined, separated by a barrier from
another deeper well. When a photon arrives at the detector an electron may
make a transition, via a state of the continuum, to the region of deeper
potential where it starts moving, this giving rise to an electric current
which is amplified by the action of the field. The important point is that,
if the external electric field is too weak the amplification does not take
place but if it is too strong some counts may be produced due to an electron
crossing the potential barrier by tunnel effect. That is, a trade-off exists
between increasing the efficiency (decreasing the false negative results)
and decreasing the dark rate (the false positive counts). I am not in a
position to prove that this fundamental trade-off is enough to prevent the
existence of the optical photon counters required for loophole-free tests of
a Bell inequality. However, it is equaly difficult to prove rigorously that
there are no fundamental constraints preventing optical photon counters
reliable enough to allow loophole-free tests of local realism.

The current wisdom that the difficulties for manufacturing reliable optical
photon counters are not fundamental derives from a theoretical prejudice,
namely that optical photons are particles like electrons or atoms. If this
were true there would be no reason why detectors could not be manufactured
having 100\% efficiency and low noise. But I think that it is closer to the
truth the assumption that photons are just quanta of the electromagnetic
field, but not particles (Willis Lamb has supported strongly this opinion%
\cite{Lamb}). There are two arguments, at least, against optical photons
being particles similar to electrons or atoms. Firstly there is no position
operator for photons in quantum mechanics, and secondly the photon number is
usually not well defined. That is, common states of light, like laser light
or thermal light, have an indefinite number of photons. A photon is (or
should be associated to) a wavepacket in the form of a needle whose length
is of the order of the coherence length, which for atomic emissions means
centimeters, and several wavelengths in transverse dimensions. This
associates a volume bigger than 10$^{16}$ atomic volumes to a typical
optical photon. In sharp contrast, a gamma ray photon may be associated to a
volume smaller than that of an atom. If we take the atomic volume as
standard, we are led to say that high energy photons are localized entities
(behaving mainly as particles) whilst optical photons are not localized
(behaving mainly as waves).

In any test of local realism using photons, it is necessary to measure both,
the position of the photon and another quantity like polarization or phase.
The former may be called a particle property whilst the latter is a wave
property. Thus, if we remember the Bell inequality $\left( \ref{BI1}\right)
, $ it is natural to associate the parameters $\zeta $ (detection
efficiency) and V (visibility of the polarization correlation curve) to
those two quantities and conclude that the Bell inequality forbids a photon
behaving as a particle and as a wave at the same time. In contrast, the
tested inequality $\left( \ref{CS}\right) $ just constrains the ``amount of
wave behaviour, V ''. Thus its violation means that we cannot dismiss the
wave character of optical photons. On the other hand, tests using gamma rays
do not have any problem with the position measurement (i. e. the efficiency
of detection), but there are difficulties for a precise measurement of
polarization, as commented in section 6. Thus I propose that, in tests using
photons, a trade-off exists between measurability of position and
measurability of polarization, trade-off quantified by the Bell inequality $%
\left( \ref{BI1}\right) $. The ``corpuscular'' property (position), may be
accurately measured only in photons much smaller than atoms, like gammas,
the ``wave'' property (polarization), in those much larger than atoms, like
optical photons.

These and other examples suggest that quantum mechanics may be compatible
with\textit{\ }local realism, the violations predicted deriving from the
idealizations used in the standard calculations, like perturbative
approximations, neglect of tunneling, etc. Maybe the reader does not agree,
but certainly the problem is open and the sober attitude is to analyze the
empirical results without the common bias that the validity of local realism
would imply that quantum theory is wrong.

\section{Digression on philosophy and sociology of science}

For the analysis of significance of the results obtained in the performed
tests of local realism it is convenient to make a digression on philosophy
and sociology of science. The pragmatic approach to quantum mechanics, which
is the basis of the Copenhagen interpretation, has led to an
``antimetaphysical'' attitude, that is the idea that science should not be
constrained at all by any philosophical principle. I think that this
position is not correct. Of course, the philosophy of the natural world
should rest upon knowledge derived from science, and not viceversa, but it
is also true that science itself rests upon some philosophical principles.

One of the central principles of the philosophy of science is that there is
not symmetry between confirmation and refutation of a theory. In fact,
although a single experiment may refute a theory, no theory can ever be
absolutely confirmed by experiments, a principle stressed by Karl Popper\cite
{Popper}. Thus the only possibility to increase the degree of confidence in
a theory is to perform many experiments able to refute it. If the results of
these experiments are compatible with the theory, it becomes increasingly
supported. We may apply this philosophy to the tests of Bell\'{}s
inequalities. As more time elapses without a loophole-free violation of
local realism, greater should be our confidence on the validity of this
principle.

Another philosophical point which is required in any serious discussion of
the present status of local realism is that established theories are
protected, a fact stressed by Imre Lakatos\cite{Lakatos}. That is, when a
new discovery seems to contradict the theory, it is always possible to
introduce some auxiliary hypotheses which allow interpreting the new finding
within the accepted theory. It is well known the example put by Lakatos on
the hypothetical observation of an anomaly in the motion of a planet. It
could be explained, without rejecting Newton's gravitational theory, by the
existence of another, unknown, planet. If this is not found by observation
in the predicted place, it might be assumed that there are two planets
instead of one, etc. Indeed, it is a historical fact that no theory has been
rejected by its contradiction with a single or even several experiments (e.
g. Newton\'{}s gravity by the anomaly in the motion of Mercury). The theory
survives until a new, superior, theory is available ( e. g. Newton's gravity
survived until the appearance of general relativity.) The consequence of
this sociological fact is that any argument \textit{for} a established
theory is accepted without too much discussion, but any argument \textit{%
against} the theory is carefully analyzed in order to discover a flaw. Thus,
even a honest experimentalist will devote much more care searching for
possible errors if an experiment contradicts the assumed predictions of
quantum mechanics than if it confirms the theory.

A good example of this behaviour has happened in the early, atomic-cascade,
tests of Bell's inequalities. As said in section 5 the first experiment of
that kind was performed by Freedman and Clauser\cite{Freedman} and the
results agreed with quantum predictions. The second experiment was made by
Holt and Pipkin (see, e.g. the reviews by Clauser and Shimony\cite{CS} or by
Duncan and Kleinpoppen\cite{Selleri6}.) The results of the experiment
disagreed with quantum predictions but did not violate the inequality $%
\left( \ref{FC}\right) $ tested. The consequence is that the experimental
results were never formally published and many people (including the
authors) made a careful search for possible sources of error. The
Holt-Pipkin experiment had two main differences with the Freedman-Clauser
one: 1) the use of a cascade of atomic mercury, instead of calcium, and 2)
the use of calcite polarizers, instead of polarizers made of piles of
plates. In order to clarify the anomaly, Clauser\cite{Clauser} ``repeated''
the Holt-Pipkin experiment, that is performed a new experiment using mercury
but, again, piles of plates as polarizers. This time the results agreed with
quantum predictions and violated the tested inequality $\left( \ref{FC}%
\right) $. However the use of calcite may be very relevant because it has an
extremely good extinction ratio, less than 10$^{-4}$ to be compared with
0.02 for typical piles of plates. In contrast calcite possesses bad
efficiency for maximum transmission of linear polarized light, about 80\% to
be compared with 98\% for typical piles of plates. But there are arguments
supporting the opinion that it is the minimal, and not the maximal,
transmission of the polarizer what matters\cite{MS89}. In spite of this
fact, the Holt-Pipkin experiment has never been repeated in the sense of
using calcite polarizers.

\section{Present status of local realism at the empirical level}

Now we arrive at the crucial question: Is local realism a valid principle of
physics?. The current wisdom is that it has been definitely refuted by the
optical experiments already performed, modulo some loopholes due to
nonidealities which, it is added, are quite common in experimental physics.
But, as explained in section 5, this is not true for the atomic-cascade
experiments (e. g. Aspect\'{}s) because they do not discriminate between
local realism and quantum mechanics, not evedn in the ideal case. We are
left with experimental tests involving optical photons produced in the
process of parametric down-conversion (e.g. the mentioned experiment by
Kurtsiefer et al\cite{Kurt}). As discussed in section 5, these experiments
cannot tests (genuine) Bell inequalities due to the lack of reliable photon
counters. If we exclude the down-conversion experiments, the evidence
against local realism is meager because all other tests present greater
difficulties. It is true that the efficiency loophole has been closed in
experiments with atoms\cite{Rowe}, what has been used as an argument \textit{%
against} the validity of local realism \cite{Grangier}. In my opinion the
fact that loopholes appear in every experiment is an argument \textit{for}
it. Indeed, it suggests that nature preserves local realism in every case.
Actually experiments like that of Rowe et al.\cite{Rowe} do not test local
realism but non-contextuality (see Sec. 2), something which is not a
principle to be mantained.

In any case I claim that \textit{local realism is such a fundamental
principle that should not be dismissed without extremely strong arguments}.
It is a fact that there is no direct empirical evidence at all for the
violation of local realism. The existing evidence is just that quantum
mechanical predictions are confirmed, in general, in tests of (non-genuine
Bell) inequalities like $\left( \ref{FC1}\right) $ or $\left( \ref{CHSH1}%
\right) .$ Only when this evidence is combined with theoretical arguments
(or prejudices) it might be arg\"{u}ed that local realism is refuted. But,
in my opinion, this combination is too weak for such a strong conclusion.
Thus I propose that \textit{no loophole-free experiment is possible which
violates local realism.}

This proposal remembers other negative statements, derived from failures at
the experimental level, which have been extremely important in the history
of physics. I shall put two examples. After James Watt made his heat engine
in 1765, many people attempted to increase the efficiency, but in some sense
they failed. In fact, nobody was able to make a \textit{perpetuum mobile}
(of the second kind), that is an engine able to produce useful work by just
cooling a large reservoir like the sea. It took sixty years to be realized,
by Sadi Carnot, that the aim was impossible because a (large) part of the
extracted heat should necessarily go to a colder reservoir. Carnot\'{}s
discovery led soon to the statement of one of the most important principles
of physics:\textit{\ the second law of thermodynamics}. Another example is
the question of the absolute motion of the Earth. Several attempts at
measuring it failed, the most sophisticated made by Michelson and Morley in
1887. The failure was ``explained'' less than 20 years later by Einstein
with the hypothesis that absolute motion does not exist. Again, a repeated
experimental failure led to a fundamental physical law: \textit{the
relativity principle}.

Ian Percival\cite{Percival} has pointed out that, in classical physics, the
second law of thermodynamics does not contradict the laws of (Newtonian)
mechanics, but nevertheless it restrict the possible evolutions of physical
systems. He proposed that a similar physical principle might prevent the
violation of local realism without actually contradicting quantum mechanics.
In my view this is an interesting observation, because I presume that it is
the second law, with quantum noise taken into account, what may prevent the
violation of local realism in the quantum domain. I think that a better
understanding of the laws of thermodynamics at the quantum level is
required. Indeed, the traditional interpretation of the third law (zero
entropy at zero Kelvin) seems difficult to be reconciled with the existence
of (non-thermal) quantum vacuum fluctuations. In summary, a serious
attention to the loopholes in the empirical tests of the Bell inequalities,
rather than their uncritical dismissal, may improve our understanding of
nature.

In any case the validity of local realism may be either refuted by a single
loophole-free experiment or increasingly confirmed by the passage of time
without such an experiment. This is the motivation for the title of the
present article.

\end{document}